\documentclass[aps,prl,reprint,groupedaddress,amsmath,amssymb]{revtex4-2}

\usepackage{graphicx}

\usepackage{xcolor}

\begin{document}


\title{Earth's Infrared Background:\\A Physics-Based Null Hypothesis for the Global-Scale Subannual Variability of Outgoing Longwave Radiation}


\author{Ofer~Shamir}
 \email{Contact author: ofer.shamir@courant.nyu.edu}
\author{Edwin~P.~Gerber}%
\affiliation{%
 Courant Institute of Mathematical Sciences, New York University, New York, New York 10012.
}%


\date{\today}

\begin{abstract}
Much of the Outgoing Longwave Radiation (OLR) emitted to space can be described as a noisy ``background'' of random variability. A rigorous characterization of this background provides an objective null spectrum and enables the isolation of atmospheric phenomena within OLR observations, such as waves and storms. Previously, the background has only been considered in the Tropics. Here we study the background on global, subannual scales and focus on its physical origins. We define the background as isotropic fluctuations implied by the fluctuation-dissipation theorem in response to internal atmospheric variability on small spatiotemporal scales. We use a stochastically forced energy balance climate model that generates a red spectrum in space and time, consistent with observations. By fitting the model to OLR measurements from satellites, we find that the background fluctuations have an upper bound of about 400~km and 2.5~days on their spatiotemporal (de)correlations, between meso-scale and synoptic-scale weather.
\end{abstract}


\maketitle

\clearpage

\section{Introduction \label{sec:intro}}

Only a small fraction of the Outgoing Longwave Radiation (OLR) emitted to space, about 17\%~\cite[40 out of 240~W\,m$^{-2}$,][]{trenberth2009earth, loeb2009toward}, originates directly at the surface. The remaining 83\% only make it out to space after having been absorbed and re-emitted by greenhouse gases (about 70\% of the total OLR, or 170~W\,m$^{-2}$) and clouds (about 13\%, or 30~W\,m$^{-2}$). Therefore, in addition to its role in determining the global energy budget, OLR also contains the “footprints” of atmospheric variability. On subannual time scales (Fig. 1A), OLR exhibits enhanced variability, relative to the global mean variance of 702~[W\,m$^{-2}$]$^{2}$, over the Intertropical Convergence Zone~\cite{schneider2014migrations, liu2020observed}, monsoonal regions, the Indian Ocean and the Maritime Continent, associated with the Madden–Julian oscillation~\cite{knutson1986global, weickmann1990shift, wheeler2004all, zhang2005madden, jiang2020fifty}, and the midlatitude storm track regions~\cite{shaw2016storm}. It also exhibits suppressed variability in the polar regions and in regions associated with the El Ni\~{n}o-Southern Oscillation~\cite{philander1983nino, ardanuy1986nino, chelliah1992large, chiodi2013nino, timmermann2018nino, fajary2019contributing}, but the latter are poorly resolved with a climatological record. Much of the observed OLR variability, however, is best described as random variability, which we term Earth’s Infrared Background (EIB). To identify the footprints of atmospheric phenomena quantitatively, it is important to provide a rigorous description of the background so that an objective null hypothesis can be established.

\begin{figure}
	\includegraphics{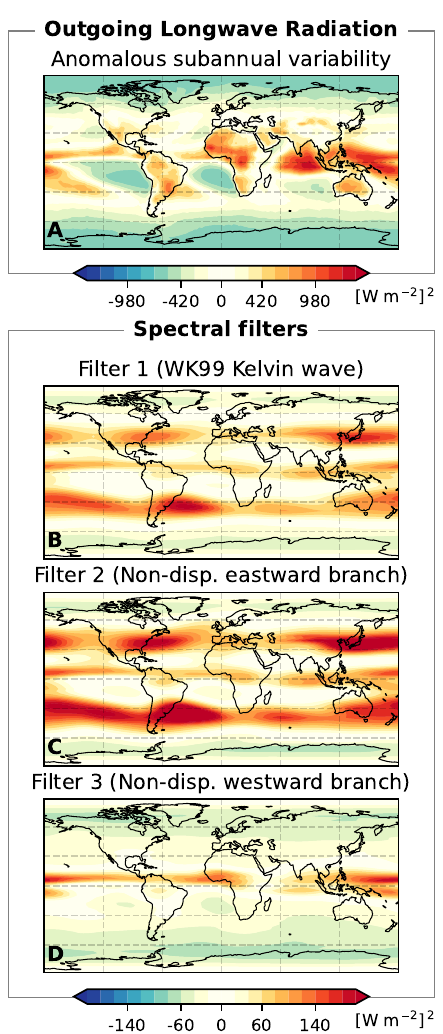}
	\caption{\label{fig:fig01}\textbf{The variance of Outgoing Longwave Radiation (OLR)}
		from satellite observations during the penultimate standard climate normal (1 January, 1981 to 31 December, 2010). \textbf{(A)} Deviation of the subannual OLR variance from the global mean (702~[W~m\textsuperscript{-2}]\textsuperscript{2}). \textbf{(B-D)} Anomalous OLR variance after filtering in spectral space using Filters 1-3 defined in Fig. 4F, and removing the background. Filter 1 in \textbf{(B)} corresponds in spectral space to the filter used by~\citet[WK99]{wheeler1999convectively} to identify Kelvin waves, but the global application reveals eastward wave activity in the midlatitude storm tracks as well. In each panel the background variance is removed. For the full OLR in panel \textbf{(A)}, the background is uniform and equal to the global spatial mean by construction.}
\end{figure}

This issue is typically encountered in the Tropics, where the identification of the tropical OLR background is essential for the quantitative analysis of equatorial waves and related coherent features~\cite{knippertz2022intricacies}. However, despite its importance, a consensual definition is still lacking, with different studies using varying approaches to estimate the background~\cite{masunaga2006madden, hendon2008some, gehne2012spectral, kikuchi2014introduction, marques2018diagnosis, kikuchi2018convectively}. Perhaps the most prevalent approach consists of successive smoothing~\cite{wheeler1999convectively}, which gradually redistributes the power in spectral space towards a uniform distribution, i.e., leading to an asymptotically white background. Alternatively, based on the observation that the OLR has a broad red background, in the sense that its power decays with decreasing spatiotemporal scales, some studies assume a priori that the background follows a red-noise process~\cite{masunaga2006madden, hendon2008some, kikuchi2014introduction, marques2018diagnosis}, albeit only in time. The most relevant approach in the present context is that of stochastic modeling pioneered by~\citet{hottovy2015spatiotemporal}, where the background is modeled as the response to a white noise forcing representing turbulent eddy fluxes. Standard results in stochastic climate modeling~\cite{hasselmann1976stochastic, north2011correlation} confirm that the resulting background in this approach is indeed broad sense red, a first-order process in time and second-order in space.

In this work, we consider a global OLR background on subannual time scales, and define it as isotropic variability implied by the fluctuation-dissipation theorem~\cite[FDT,][]{kubo1966fluctuation} in response to the internal variability of the atmosphere. The underlying idea is that, under weak perturbations, the noise level of the system is determined by its dissipative properties and equilibrium distribution. Like classical Brownian motion, we assume a scale separation between the response and the forcing, such that the latter can be approximated as a white noise process. Unlike classical Brownian motion, however, the equilibrium distribution is not derived from first principles and is only estimated empirically. In addition, the dissipation is modeled using the ``simplest'' mathematical terms, and the associated coefficients are also estimated empirically to fit observations. Although the resulting background is empirical, it meets our aim of providing an objective and quantitative null hypothesis for isolating atmospheric variability on subannual time scales. In addition, while employing a stochastic model similar to that of~\citet{hottovy2015spatiotemporal}, the present approach places an additional constraint on the power spectrum of the forcing.

The assumption that the background is statistically isotropic is the spherical equivalent of statistical homogeneity in time series analysis and provides the most conservative null hypothesis for the variance. This amounts to the assumption that any anisotropy observed in the data reflects genuine physical variability rather than an artifact of the noise. We stress that the observed OLR is far from isotropic, as evident by its non-uniform variance in Fig. 1A (a necessary condition for statistical isotropy). However, the identified regions of enhanced/suppressed variability in Fig.1A are associated with non-equilibrium phenomena and are the very things the null hypothesis is designed to detect.


\begin{figure*}
	\includegraphics[width=0.9\textwidth]{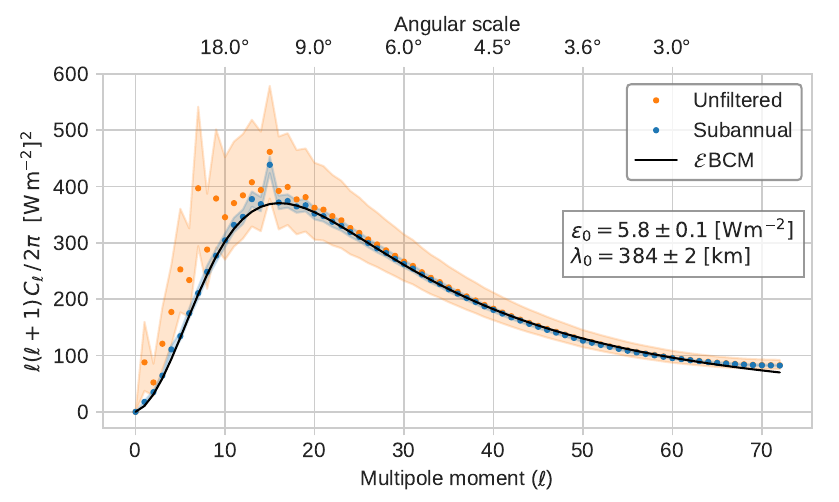}
	\caption{\label{fig:fig02}\textbf{The Angular power spectrum} is obtained by averaging OLR fluctuations in spectral space, $F_{lm}^{}(t) F_{lm}^{*}(t)$, over time and $m$ (with $|m| \leq l$) to estimate the angular variance ${C}_{l} = \langle F_{lm}^{} F_{lm}^{*} \rangle$ (orange). The angular variance provided by the $\mathcal{E}$BCM (Eq.~\ref{eq:ebcm-solutions-a} with $t = t'$, black line) is fit to the data on subannual time scales (including frequencies above 1/360 cpd, blue) using nonlinear least squares to estimate $\epsilon_{0}$ and $\lambda_{0}$. With the scaling on the ordinate, the figure represent the power per logarithmic interval in angular scale. The peak at $l=15$ (12$^{\circ}$) is an artifact corresponding to the satellite swath half-width of about 1,250~km. The uncertainty in the raw data is associated with the variance across $m$, while the uncertainty in the subseasonal data is associated with the variance across both $m$ and the temporal windows (see Supporting Information S3).}
\end{figure*}

\section{A Stochastically Forced Energy Balance Climate Model ($\mathcal{E}$BCM) \label{sec:model}} 

A quantitative description of the background requires an empirically consistent model. The most parsimonious, lowest order, model of isotropic fluctuations that captures the above physical picture and the observed OLR variance on subannual time scales is given by the following stochastic differential equation:
\begin{equation} \label{eq:ebcm}
    \tau_{0} \frac{\partial F}{\partial t} - \lambda_{0}^{2} \nabla^{2} F + F = S,
\end{equation}
where $F$ represents OLR fluctuations as a function of time, longitude, and latitude, and $S$ represents a stochastic forcing due to internal fluctuations of the atmosphere. The model has a total of 3 free parameters, a time scale $\tau_{0}$, which sets the temporal decorrelation, a length scale $\lambda_{0}$, which sets the spatial decorrelation, and an amplitude $\epsilon_{0}$, which sets the variance and enters through the forcing $S$ as described below.

This model has been extensively analyzed in the context of energy balance climate models~\cite[$\mathcal{E}$BCMs,][]{north1975theory,north1981energy}, where some of its applications include the study of second-order surface temperature statistics~\cite{kim1991surface,north2011correlation}, and climate predictability bounds~\cite{north1981predictability}. Save for the scale-dependence of the forcing, described below, the present analysis follows these works. However, while resulting in a similar model, our motivation is different. In particular, we have no reason to assume that purely random OLR fluctuations are subject to Fickian diffusion. Rather, in the present context, the Laplacian on the left-hand side is the lowest order differential operator (other than the identity) that is invariant under rotation, which guarantees that the response to a statistically isotropic forcing remains isotropic~\cite{north1981predictability, north2011correlation}, while the relaxation term guarantees that the response remains statistically stationary~\cite{hasselmann1976stochastic, frankignoul1977stochastic}. A complementary viewpoint is provided in~\citet{hottovy2015spatiotemporal}, who studied a similar model in the context of the tropical background, and suggested that it represents the effects of mean and eddy flux convergence in a turbulent atmosphere.

We take the forcing to be Gaussian white noise in time and statistically isotropic in space. The former implies that the forcing is $\delta$-correlated in time, assuming a priori that the forcing decorrelates much faster than the response. The latter implies that the forcing is $\delta$-correlated in Spherical Harmonic space, and depends at most on the multiple moment~\cite{obukhov1947statistically}, i.e.,
\begin{equation} \label{eq:s-statistics}
    \langle S_{lm}^{}(t) S_{l'm'}^{*}(t') \rangle = 2 \epsilon_{0}^{2} \tau_{l} \delta_{ll'} \delta_{mm'} \delta(t-t'),
\end{equation}
where $S_{lm}$ are the spectral coefficients of $S$ of degree (total wavenumber) $l$ and order (planetary zonal wavenumber) $m$, asterisks denote complex conjugates, angle brackets denote ensemble averages, sub-scripted $\delta_{ij}$ is the Kronecker delta, and argumented $\delta()$ is the Dirac delta. In a key departure from earlier works~\cite{north2011correlation, hottovy2015spatiotemporal}, we find it necessary to allow the forcing to be scale dependent ($l$ dependent) to explain the observed OLR. Here $\tau_{l} = \tau_{0} / [1 + \lambda_{0}^{2} l (l+1) / a^{2}]$, where $a$ is the mean radius of the Earth, which is the only choice consistent with the FDT (Supporting Information S2). We note that the spatial decorrelation of the forcing is much faster than that of the resulting response (Supporting Information S3), so that the two are well separated in both space and time.

The process in Eq.~(\ref{eq:ebcm}) is fully determined by its covariance function, which for an isotropic process depends only on the scale $l$ and is termed angular covariance. The asymptotic angular covariance, at times much greater than the decorrelation time ($t, t' \gg \tau_{l}$), is~\cite[e.g.,][]{zwanzig2001nonequilibrium}:
\begin{equation}  \label{eq:ebcm-solutions-a}
    C_{l}(t, t') = \langle F_{lm}^{}(t) F_{lm}^{*}(t') \rangle = \epsilon_{l}^{2} e^{-|t-t'| / \tau_{l}},
\end{equation}
where $F_{lm}$ are the spectral coefficients of $F$ and $\epsilon_{l} = \epsilon_{0} \tau_{l} / \tau_{0}$. For the background spectrum, we will use the corresponding power spectral density (PSD):
\begin{equation}  \label{eq:ebcm-solutions-b}
    \hat{C}_{l}(\omega) = 2 \epsilon_{l}^{2} \tau_{l} / \left[\omega^{2} \tau_{l}^{2} + 1\right],
\end{equation}
where hats denote the Fourier modes in the frequency domain, obtained from the covariance via the Wiener–Khinchin theorem, i.e.,
\begin{equation*}
\hat{C}_{l}(\omega) = \int_{-\infty}^{\infty} C_{l}(\tau) e^{-i\omega\tau} d\tau.
\end{equation*}
%

\section{Estimating the model parameters from observations}

\begin{figure*}
	\includegraphics[width=0.9\textwidth]{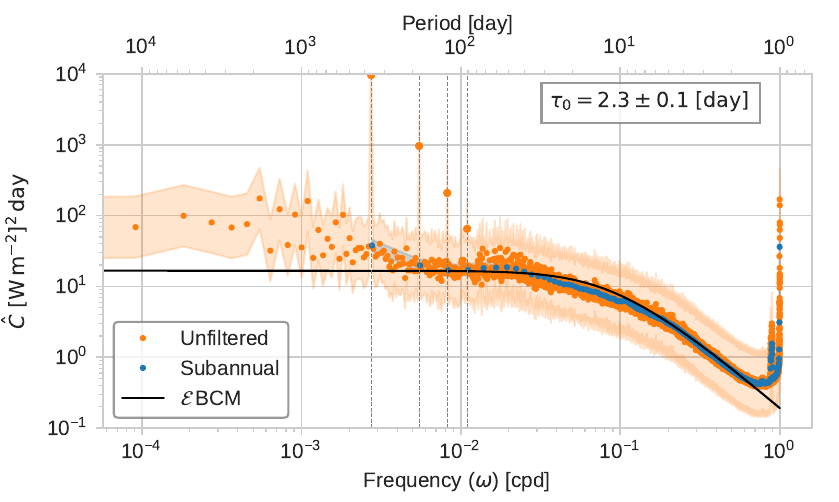}
	\caption{\label{fig:fig03}\textbf{The temporal power spectrum} is obtained by applying a discrete Fourier transform to the time-dependent Spherical Harmonic coefficient to estimate the PSD $\tilde{F}_{lm}^{}(\omega) \tilde{F}_{lm}^{*}(\omega) / \Delta \omega$, where $\Delta \omega$ is the frequency resolution, and averaging over $m$ (with $|m| \le l$) and $l$ (orange). The PSD provided by the $\mathcal{E}$BCM (Eq.~\ref{eq:ebcm-solutions-b}, black line) averaged over $l$ is fit to the data on subannual time scales (including frequencies above 1/360 cpd, blue) using the values of $\epsilon_{0}$ and $\lambda_{0}$ estimated in Fig.~\ref{fig:fig02}, and nonlinear least squares to estimate $\tau_{0}$. The figure represents the power per frequency. Vertical dashed lines indicate the annual cycle at 1/365 cpd and subsequent three harmonics: $(2,3,4)/365$~cpd. The uncertainty in the raw data is associated with the variance across $m$, while the uncertainty in the subseasonal data is associated with the variance across both $m$ and the temporal windows.}
\end{figure*}

Before estimating the background, we first use Eqs.~(\ref{eq:ebcm-solutions-a}) and (\ref{eq:ebcm-solutions-b}) to estimate $\epsilon_{0}$, $\lambda_{0}$, and $\tau_{0}$ from observations (Supporting Information S1) and discuss their physical meaning.

Consider first the angular power spectrum of the EIB (Fig.~\ref{fig:fig02}), obtained by multiplying the time-dependent Spherical Harmonic coefficient by their complex conjugates $F_{lm}^{}(t) F_{lm}^{*}(t)$ and averaging over time and $m$ (with $|m| \leq l$) to estimate the angular variance ${C}_{l} = \langle F_{lm}^{} F_{lm}^{*} \rangle$. By the Wiener-Khinchin theorem, the latter is related to the PSD via $C_{l}(t=t') = \int_{-\infty}^{\infty} \hat{C}_{l}(\omega) d\omega \, / 2\pi$, and therefore provides a measure of the angular power spectrum in terms of the frequency-averaged PSD. At small spatial scales (large $l$), the $\mathcal{E}$BCM (black line) can accurately fit the raw variance (orange) by setting $t = t'$ in Eq.~(\ref{eq:ebcm-solutions-a}) and using non-linear least squares to estimate $\epsilon_{0} = 5.8 \pm 0.1$~W\,m$^{-2}$ and $\lambda_{0} = 384 \pm 2$~km. At large spatial scales (small $l$), however, there are externally forced variations and the $\mathcal{E}$BCM can only explain the observed variance on subannual time scales with frequencies above $1/360$ cpd (blue). The peak at $l=15$ is an artifact corresponding to the satellite swath half-width of about 1,250~km. A weaker imprint of the satellite swath width is also found at $l=13$ (Supporting Information S3).

In the context of the stochastic model used here, $\lambda_{0}$ determines the spatial decorrelation of the EIB. More precisely, $\lambda_{0}$ provides an upper bound on the spatial decorrelation. The effective decorrelation decreases with increasing frequency (Supporting Information S3) and is not simply an e-folding scale~\cite{north2011correlation}. The estimated value of $\lambda_{0} = 384 \pm 2$~km is well below the Nyquist wavelength of 5$^{\circ}$ in the data, consistent with random fluctuations with spatial decorrelation smaller than the smallest resolvable waves. In addition, $\lambda_{0}$ is also smaller than the typical Rossby deformation radius, providing further reassurance that the background is associated with random fluctuations and suggesting that it is constrained by linear wave dynamics.

Next, consider the temporal power spectrum (Fig.~\ref{fig:fig03}), obtained by applying a discrete Fourier transform on the time-dependent Spherical Harmonic coefficient to estimate the PSD $\tilde{F}_{lm}^{}(\omega) \tilde{F}_{lm}^{*}(\omega) / \Delta \omega$, where $\Delta \omega$ is the frequency resolution, and averaging over $m$ (with $|m| \le l$) and $l$. The uptick at $\omega = 1$ cpd (the Nyquist frequency) likely includes contributions from the diurnal cycle in clouds, rainfall, and stratospheric tides, though these contributions are obscured by aliasing from higher-frequency variability. The uptick at $\omega = 0.9$ cpd (a period of $1.1$~day) is associated with the satellite orbit~\cite{wheeler1999convectively}, and the upward inflection at $\omega >0.6$ cpd is the result of spectral leakage (Supporting Information S1). The dominant harmonics at $(1,2,3,4)/365$~cpd (vertical dashed lines) correspond to the annual, semi-annual, and seasonal cycles, and were explicitly removed before fitting the model (Supporting Information S1). Having estimated $\epsilon_{0}$ and $\lambda_{0}$ from the angular power spectrum, $\tau_{0}$ can now be estimated by averaging Eq.~(\ref{eq:ebcm-solutions-b}) over $l$ and using non-linear least squares (black line), yielding $\tau_{0} = 2.3 \pm 0.1$~days. Consistent with the angular variance in Fig.~\ref{fig:fig02}, it is also evident in this figure that the $\mathcal{E}$BCM fits the temporal power spectrum only on subannual time scales with frequencies above 1/360 cpd.

In the context of the stochastic model used here, $\tau_{0}$ provides an upper bound on the temporal decorrelation. The estimated value is similar to those obtained in the Tropics by~\citet{hendon2008some} assuming a red noise process in time, between 3 days at low zonal wavenumbers and 1 at high wavenumber. It is also in the ballpark of the 4 day decorrelation time used in the stochastic model of~\citet{hottovy2015spatiotemporal}. While not strictly comparable, the common insight that stems from both these works in the Tropics and the present global scale analysis is that OLR decorrelates on intermediate time scales, between fast convection at small spatial scales and slower linear damping at large scales~\cite{Holton1972, Sardeshmukh1984, Lin2008, Romps2014, shamir2023matsuno}. The effective temporal decorrelation has an $l$-dependent e-folding scale $\tau_{l} = \tau_{0} / [1 + \lambda_{0}^{2} l (l+1) / a^{2}]$, as implied by Eq.~(\ref{eq:ebcm-solutions-a}). At $l=0$, the temporal decorrelation is at the lower limit of synoptic-scale weather, typically taken as 2~days. At $l=20$ (about 1000~km), the decorrelation is 22~hours, at the upper limit of meso-scale weather, typically taken as 1~days and 1000~km. Finally, at $l=72$ (about 300~km), the decorrelation is 3~hours, at the lower limit of meso-scale weather. This last value is important since, by assumption, the temporal correlation of the forcing is much shorter than that of the response. Therefore, the forcing mechanism(s) must decorrelate faster than meso-scale weather, to allow for sufficient scale separation between the two.


\section{Estimating the background spectrum and filtering in spectral space \label{sec:space-time-analysis}}

\begin{figure*}
	\includegraphics[width=0.9\textwidth]{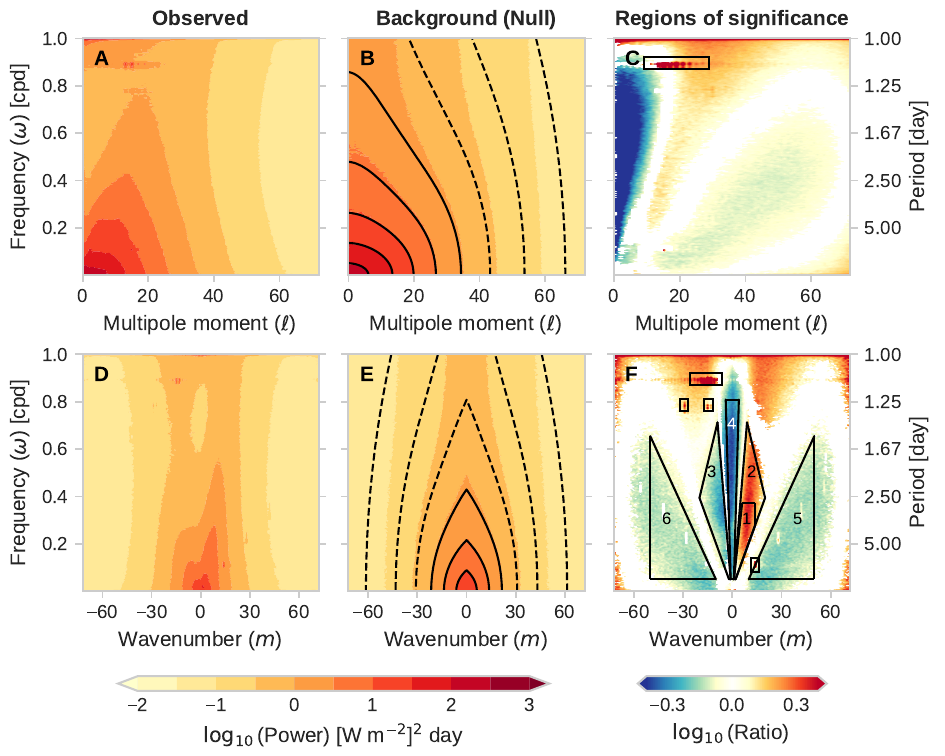}
	\caption{\label{fig:fig04}\textbf{Space-time spectra.}
		The power spectral density (PSD) of the observed OLR are obtained by averaging $\tilde{F}_{lm}^{}(\omega) \tilde{F}_{lm}^{*}(\omega) / \Delta \omega$, where tildes denote the Discrete Fourier Transform and $\Delta \omega$ is the frequency resolution, over zonal wavenumber $m$ (\textbf{A}) and total wavenumber $l$ (\textbf{D}) with $|m| \leq l$. (\textbf{B, E}) The corresponding PSD of a random realization of the background, generated by solving the spectral space version of Eq.~(\ref{eq:ebcm}) as an $l$-dependent Ornstein--Uhlenbeck process (Supporting Information S1), and having the same sample size and sample rate as in the observations (\textbf{A, D}). For comparison, black contours mark the analytic PSD given by Eq.~(\ref{eq:ebcm-solutions-b}), highlighting uncertainty associated with finite sampling. (\textbf{C, F}) Regions of significance, where the ratio of observed to background PSD is significantly different from 1 (Supporting Information S1). With the logarithmic scaling, the power in panel C (F) is simply the difference between the spectra in panels A and B (D and E). The spectral filters used in Fig. 1(B-D) and in Supporting Information S3 are outlined in panel (\textbf{F}) and labeled 1-6. Black rectangles in panels (\textbf{C, F}) indicate satellite artifacts.}
\end{figure*}

The full description of the EIB consists of its joint power distribution in $\omega, l, m$ space. We consider the PSD obtained by first applying a discrete Fourier transform to the time-dependent Spherical Harmonic coefficients to estimate $\tilde{F}_{lm}^{}(\omega) \tilde{F}_{lm}^{*}(\omega) / \Delta \omega$, and then averaging over $m$ (with $|m| \le l$) or $l$ (with $|m| \le l \le 72$) to obtain the PSD as a function of $\omega$ and $l$ (Fig. 4, top row) or $\omega$ and $m$ (Fig. 4, bottom row), respectively. For the latter, the background (Fig. 4E) depends on $m$ implicitly through the $m$-dependent averaging, although the $\mathcal{E}$BCM does not.

Having estimated the model parameters, we generate a random realization of the background (Fig. 4B,E) by solving the spectral space version of Eq.~(\ref{eq:ebcm}) as an $l$-dependent Ornstein--Uhlenbeck process (Supporting Information S1), having the same sample size and sample rate as the observed OLR. The advantage of this approach is that the realized background accounts for the effects of finite sampling, therefore providing a better basis for comparison than the analytic PSD (black contours) given by Eq.~(\ref{eq:ebcm-solutions-b}). Significant signals (Fig. 4C,F) are then identified as $\omega, l, m$ combinations where the ratio of observed to realized PSDs is significantly different from one (see Supporting Information S1 for statistical analysis). In other words, the null hypothesis is that the power follows a red‑noise distribution in both space and time, and the alternative hypothesis is that it differs.

Aside from satellite artifacts (black rectangles) and spectral leakage at high frequencies (above 0.8 cpd; see also Supporting Information S1), the global background reveals distinct regions of significance in Fig. 4F. In order to relate these regions to known atmospheric variability, at least in part, we examine below the resulting spatial variance after filtering in spectral space according to these regions. The chosen filters are outlined by black (pseudo-) triangles and labeled 1-6. While the background is isotropic by definition, the application of the filters introduces anisotropies. Therefore, the filters were applied to both the observed OLR and the realized background, and their spatial variance were subtracted.

\sloppy For comparison, Filter 1 corresponds to the exact region of spectral space used by~\citet{wheeler1999convectively} to filter Kelvin waves in the Tropics, but it is applied to the global subannual variability rather than only equatorially symmetric summer variability. The resulting spatial variance is shown in Fig. 1B. Variability associated with equatorial Kelvin waves can indeed be identified in the Tropics. However, when applied globally, the variance corresponding to this region of spectral space also captures variability in the midlatitude storm track regions, which is closely associated with midlatitude baroclinic Rossby waves~\cite{shaw2016storm}. 

Filter 1 makes up part of a larger region of significance identified in Figure 4F and delineated by Filter 2, which represents non-dispersive eastward propagating variability. The resulting spatial variance (Fig. 1C) is generally similar to that of Filter 1, but with stronger variability consistent with the larger area in spectral space, which is, in turn, consistent with the broader spectrum of midlatitude baroclinic Rossby waves toward smaller scales and higher frequencies. 

Filter 3 is a mirror image of Filter 2 about wavenumber 0 and represents non-dispersive, westward propagating variability. The variability in the midlatitude storm track regions is clearly suppressed compared to the eastward propagating branch in Filter 2 (Fig. 1D), and the polar regions are also suppressed compared to the background. Enhanced spatial variance is focused only in the Tropics. Its spatial pattern bears some resemblance to that of the westward propagating equatorial Rossby wave~\cite[e.g.,][see also Supporting Information S3]{wheeler1999convectively}, despite covering a different region of spectral space. Even in the Tropics, it exhibits less variability than Filter 2, despite covering an area of the same size in spectral space. 

Together, the picture that stems from Filters 2 and 3 is consistent with traditional midlatitude Rossby wave dynamics, not just in terms of spatiotemporal scales and geographical projection, but also in dynamical terms. The enhanced variability of the eastward branch and the suppressed variability of the westward branch are consistent with Doppler shifting by the mean winds of the extratropical jets. From this perspective, the remanent tropical pattern in Filter 3 is consistent with the weak easterly mean winds in that region. Moreover, the fact that this variability appears as non-dispersive waves in spectral space is consistent with the dispersion relation of Rossby waves in the limit of large wavelengths compared to the Rossby radius of deformation. 

Consistent with large spatial scales centered around wavenumber 0, the spatial variance corresponding to Filter 4 (Supporting Information S3) consists of zonally symmetric equator-to-pole gradient, highlighting the lower OLR variability at the poles observed in Fig. 1A. Finally, the spatial variance corresponding to Filters 5 and 6 (Supporting Information S3) highlights regional variability in the monsoonal regions and the ITCZ, in addition to the storm track regions. Additional filters are considered in Supporting Information S3 to motivate further analysis beyond the scope of this work.


\section{Conclusions and Discussion} \label{sec:discussion}

We define the Earth's Infrared Background (EIB) as isotropic, random fluctuations implied by the fluctuation-dissipation theorem (FDT), in response to internal variability of the atmosphere on small spatiotemporal scales. A simple stochastically forced energy balance climate model is capable of explaining the observed space/time spectra on subannual time scales, provided the forcing has scale dependence. The resulting (joint) space-time spectrum of the background captures the total observed variance, by construction, and is distributed in space and time according to a red noise process. The deviations from this red background reveal some atmospheric variability of interest, including equatorial Kelvin waves waves, midlatitude Rossby waves, monsoonal circulation, and the large-scale meridional circulation.

As defined here and in previous work, the background captures the same total variance as the raw data. This widely adopted terminology has led to confusion and criticism, as the word background implies a baseline level. Semantically, it is reasonable to expect a lower bound. While we have adopted the preceding terminology, a more accurate term to describe the target of our analysis would be a ``null spectrum'', which need not be a lower bound. It is based on the null hypothesis that there is no spatiotemporal structure in the data. By construction, it captures the total observed variance, but is distributed in space and time according to an idealized red noise process. Therefore, it is the distribution of power in spectral space, not the total power, that distinguishes the observed OLR from the background and enables the identification of significant signals.

If the background is restricted to be a lower bound, it implicitly assumes that significant signals consist only of enhanced variability. In contrast, from the point of view of a random null hypothesis, both enhanced and suppressed variability are possible. Physically, the variability at certain wavenumbers and frequencies can be suppressed due to anti-resonance. In the Tropics, for example, OLR was found to be suppressed as a result of anti-resonance between (dry) wave modes and (moist) convection~\cite{stechmann2017unified}. On a global scale (the present work), suppressed variability in the non-dispersive westward branch is important for completing the physical picture of the midlatitude Rossby wave dynamics in the presence of westerly jets. 

An important limitation of our null spectrum that should be kept in mind when interpreting the results is the effects of spatial variations in the time mean wind. In the Tropics, for example, the PSD of convectively coupled equatorial waves was found to be systematically modulated as a result of Doppler shifting by the varying mean wind in different tropical sectors~\cite{dias2014influence}. Moreover, much of the spread in spectral space (interpreted as a spread in the ``equivalent depth'') was attributed to this effect. It is thus possible that the regions of significance identified in the present work are ``smeared out'' by the varying mean wind across the globe. Nonetheless, the EIB reveals distinct regions of variability in spectral space associated with the dynamics of the atmosphere. 

Finally, some more interesting questions remain. Although grounded in the fundamental principle of the FDT, the forcing and damping terms were chosen for simplicity and empirical agreement, rather than derived from first principles. The question is therefore what are the relevant physical processes at play. A related question is what determines the (de)correlation scales of the background.


\begin{large}
    \begin{center}
        Acknowledgments
    \end{center}
\end{large}

\begin{acknowledgments}
O.~S. and E.~P.~G. were supported by Schmidt Sciences LLC. E.~P.~G. was also supported by the National Science Foundation through award OAC-2004572. We also acknowledge high-performance computing support from the Derecho system (doi:10.5065/qx9a-pg09) and the Cheyenne system (doi:10.5065/D6RX99HX) provided by the National Center for Atmospheric Research (NCAR), sponsored by the National Science Foundation. 
\end{acknowledgments}



\section{Supporting Information}

\renewcommand{\theequation}{S\arabic{equation}}
\setcounter{equation}{0}
\renewcommand{\thefigure}{S\arabic{figure}}
\setcounter{figure}{0}

\subsection{S1: Data and Methods \label{si:data}}
\subsubsection{Observations \label{si:obs}}
Satellite observations of Outgoing Longwave Radiation (OLR) were sourced from the Physical Sciences Laboratory of the National Oceanic and Atmospheric Administration. This product is interpolated in time and space, as described in~\citet{liebmann1996description}, to yield twice daily estimates on a 2.5$^{\circ}$ by 2.5$^{\circ}$ latitude-longitude grid. 

The analysis was carried out on the penultimate standard climate normal between 1981 and 2010, as defined by the World Meteorological Organization~\cite{WMO-report}. A climatological record provides a reasonable compromise between a sufficiently long time to effectively sample subannual variability, and a sufficiently short time for the record to remain stationary. Although the stationarity of such records in the presence of climate change has been increasingly questioned~\cite{milly2008stationarity,arguez2011definition}.

\subsubsection{Preprocessing and Filtering \label{si:preprocessing}}
By identifying the background with the response to the internal variability of the atmosphere on small spatiotemporal scales, we assume that it is independent of external, anisotropic, forcing associated with the insolation and surface properties. Within the framework of the energy balance climate model ($\mathcal{E}$BCM), these effects are represented as additive forcing~\cite{north1975theory, north1981energy}. By neglecting their contributions, we effectively assume that the decorrelation time of the background is short compared to their temporal variation. Some low-frequency filtering is therefore necessary to remove their contributions from the data. Empirically, we find that the $\mathcal{E}$BCM can accurately describe the OLR variability on subannual timescales, with frequencies above 1/360 cpd. However, this cutoff is somewhat arbitrary and the appropriate cutoff depends on the context.

In practice, the filtered data and corresponding power spectra were obtained using Welch’s Overlapping Segment Analysis as follows: First, the long-term mean was removed. In addition, the annual, semi-annual, and seasonal cycles (vertical dashed lines in Fig.~3 of the main text) were also removed by zeroing-out their Fourier components. The record was then divided into windows of 360~days, with a 180~day overlap, for a total of 59 windows. To minimize spectral leakage, the windows were tapered in time using a Hann window. In addition to effectively filtering the data, the advantage of this approach is that it provides a consistent estimator of the power spectral density~\cite{percival1993spectral,mudelsee2010climate}. 

The analysis was repeated with windows of 90~days, with a 30~day overlap, to remove further variability associated with the seasonal cycle. The results remain unchanged within the estimated uncertainty; the estimated parameters of the $\mathcal{E}$BCM in that case are $\epsilon_{0} = 5.7 \pm 0.1$~W\,m$^{-2}$ and $\tau_{0} = 2.5 \pm 0.1$~days.

\subsubsection{Spectral Analysis \label{si:spectral-analysis}}
Following the notation of the main text, let $F(\lambda, \phi, t)$ denote the physical (grid) space OLR fluctuations evaluated at (equally spaced) longitudes $\lambda$, latitudes $\phi$, and times. The open-source \textsc{SHTools} Python package~\cite{wieczorek2018shtools} was used to evaluate the spectral coefficients in Spherical Harmonics space as a function of time, $F_{lm}(t)$. Specifically, the OLR fluctuations were analyzed using the orthonormalized complex Spherical Harmonics (with the Condon–Shortley phase factor), such that
\begin{equation} \label{eq:sh-convention-orthogonality}
    \int_{0}^{2\pi}\int_{-\pi/2}^{\pi/2} Y_{l}^{m}(\lambda, \phi) Y_{l'}^{m'}{}^{*}(\lambda, \phi) \cos\phi \, d\phi \, d\lambda = \delta_{ll'}\delta_{mm'},
\end{equation}
where $Y_{l}^{m}$ are the Spherical Harmonics of degree $l$ and order $m$, the asterisks denote complex conjugates, and sub-scripted $\delta_{ij}$ is the Kronecker delta. To evaluate integrals in grid space (e.g., when comparing with the global mean OLR and confirming Parseval's theorem) the data were linearly interpolated to a Gauss-Legendre grid.

Having evaluated the time-dependent Spherical Harmonic coefficients, the frequency domain spectral coefficients $\tilde{F}_{lm}(\omega)$ were obtained by further applying a discrete Fourier transform over time. Throughout the text, we use hats to denote the continuous Fourier transform, used in the analytic results, and tildes to denote the discrete Fourier transform, used in the data. The convention used here for the former is 
\begin{equation} \label{eq:fourier-convention}
    f(t) = \frac{1}{2\pi} \int_{-\infty}^{\infty} \hat{f}(\omega) e^{i\omega t} d\omega,
    \quad
    \hat{f}(\omega) = \int_{-\infty}^{\infty} f(t) e^{-i\omega t} dt.
\end{equation} 

The power spectral density (PSD) was estimated by multiplying the spectral coefficients by their complex conjugates and dividing by the frequency resolution to obtain the power per frequency bin $\tilde{F}_{lm}^{}(\omega) \tilde{F}_{lm}^{*}(\omega) / \Delta\omega$, in units of [W\,m$^{-2}$]$^{2}$~day. To compensate for the power attenuation introduced by the Hann window (see above), the PSD was further multiplied by $8/3$. Alternatively, the frequency resolution can be replaced with the equivalent noise bandwidth.

\subsubsection{Variance Calculations \label{si:variance}}

\begin{figure*}
	\includegraphics[width=12cm]{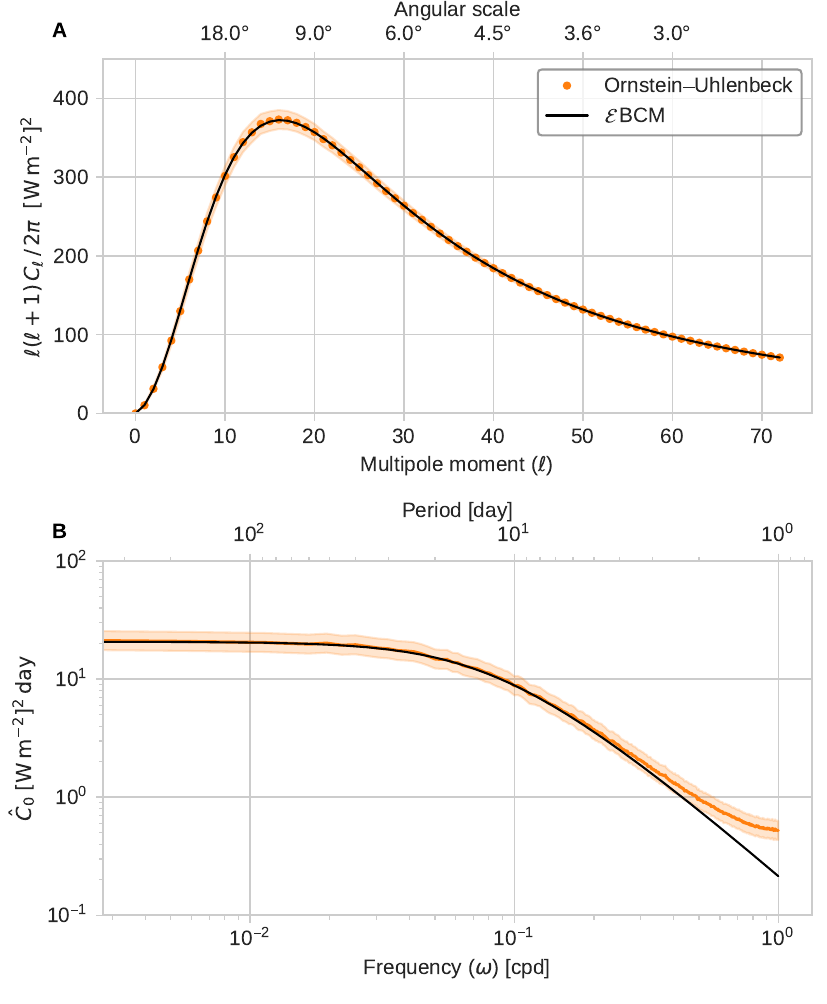}
	\caption{\label{fig:figs01}\textbf{Realized space/time spectra.}
		Same as Figures 2 and 3 of the main text, but for a random realization of the $\mathcal{E}$BCM, obtained by solving its spectral space version Eq.~(\ref{eq:ebcm-lm}) as an $l$-dependent Ornstein--Uhlenbeck process (orange). For comparison, the power spectra estimated from OLR observations is also shown (black lines, same as in Figs. 2-3 of the main text). The figure confirms that the realized background has the desired power spectra, and provides a sense of the effects of spectral leakage associated with the analysis and the finite sampling.}
\end{figure*}

By construction, the background captures the total observed OLR variance. In this section, we provide the relevant formulae and implementation details.

By definition, the global mean variance in grid space is
\begin{equation} \label{eq:global-variance-physical-space}
    \frac{1}{4 \pi} \int_{0}^{2\pi} \int_{-\pi/2}^{\pi/2} \langle F(\lambda, \phi, t) F^{*}(\lambda, \phi, t) \rangle \cos\phi \, d\phi \, d\lambda.
\end{equation}
For each window, the grid-point variance was estimated by averaging $F^{}(\lambda, \phi, t)F^{*}(\lambda, \phi, t)$ over time. The meridional integral was estimated using Gauss-Legendre quadrature. To this end, the resulting variance was first linearly interpolated in latitude from the regular grid to a Gauss-Legendre grid. The zonal integral was estimated using the trapezoid rule at the equi-distanced longitudes. The resulting global mean variance of the observed OLR, averaged over all windows, is $26.4 \pm 0.2$~W\,m$^{-2}$, where the uncertainty is the one associated with the standard error over the samples (windows).

Assuming $F$ is statistically isotropic, the corresponding global mean variance in spectral space is
\begin{equation}
    \frac{1}{4 \pi} \sum_{l=0}^{\infty} \sum_{m=-l}^{l} \langle F_{lm}^{} F_{lm}^{*} \rangle.
\end{equation}
Using Eq.~(3) with $t=t'$, the global mean variance of the $\mathcal{E}$BCM is then
\begin{equation} \label{eq:global-variance-spectral-space}
    \sum_{l=0}^{72} \frac{(2l+1)}{4 \pi} \epsilon_{l}^{2},
\end{equation}
where the summation was truncated according to the truncation order of the data. Using the estimated values of $\epsilon_{0} = 5.8 \pm 0.1$~W\,m$^{-2}$ and $\lambda_{0} = 384 \pm 2$~km to obtain $\epsilon_{l} = \epsilon_{0} / [1 + \lambda_{0}^2 l(l+1) / a^2]$, the global mean STD of the $\mathcal{E}$BCM is $26.6 \pm 0.5$~W\,m$^{-2}$, where the uncertainty is the one implied by those of $\epsilon_{0}$ and $\lambda_{0}$ assuming the two are uncorrelated.

The total variance is best described in relation to Parseval's theorem. With the above conventions for the Spherical Harmonics and the Fourier transform, the continuous form of Parseval's theorem is
\begin{eqnarray} \label{eq:parseval}
    \int_{-\infty}^{\infty} \int_{0}^{2\pi} \int_{-\pi/2}^{\pi/2} F^{} F^{*} \cos\phi \, d\phi \, d\lambda \, dt
    = \nonumber \\
    \frac{1}{2 \pi} \int_{-\infty}^{\infty} d\omega \, \sum_{l=0}^{\infty} \sum_{m=-l}^{l} \hat{F}_{lm}^{}(\omega) \hat{F}_{lm}^{*}(\omega).
\end{eqnarray}
Using the PSD in Eq.~(4) to substitute on the RHS, the total variance in the $\mathcal{E}$BCM is then
\begin{equation} \label{eq:ebcm-total-variance}
    \sum_{l=0}^{72} (2l+1) \epsilon_{l}^{2}.
\end{equation}
It can be seen that the total variance in the $\mathcal{E}$BCM is simply proportional to the global mean variance in Eq.~(\ref{eq:global-variance-spectral-space}), which is another manifestation of the fact that the angular variance is proportional to the frequency-averaged PSD. Using the estimated values of $\epsilon_{0}$ and $\lambda_{0}$, the total variance in the $\mathcal{E}$BCM is $8893 \pm 314$~[W\,m$^{-2}$]$^{2}$.

The total variance in the data was calculated using the discrete form of Parseval's theorem
\begin{eqnarray} \label{eq:parseval-discrete}
    \sum_{n=0}^{719} & & \int_{0}^{2\pi} \int_{-\pi/2}^{\pi/2} F_{n}^{} F_{n}^{*} \cos\phi \, d\phi \, d\lambda \,  
    = \nonumber \\
    & & \frac{1}{720} \sum_{n=0}^{719} \, \sum_{l=0}^{72} \sum_{m=-l}^{l} \tilde{F}_{lm}^{}(\omega_{n}) \tilde{F}_{lm}^{*}(\omega_{n}),
\end{eqnarray}
where the series were truncated according to the truncation order of the data and each temporal window has 360~days with 2 samples per day for a total of 720 samples. The left-hand side was estimated as described above for the global mean variance, yielding a total variance of $9062 \pm 308$~[W\,m$^{-2}$]$^{2}$. The right-hand side was calculated by straightforward addition, yielding a total variance of $9164 \pm 310$~[W\,m$^{-2}$]$^{2}$. In the main text, the average value of the two estimates is reported ($9113 \pm 309$~[W\,m$^{-2}$]$^{2}$).

\subsubsection{Random realizations \label{si:realizations}}

In order to generate spatiotemporal realizations of the $\mathcal{E}$BCM, we solve its spectral space version (\ref{eq:ebcm-lm}, below) as an $l$-dependent Ornstein--Uhlenbeck (OU) process. Consider a generic OU process $X_{t}$:
\begin{equation} \label{eq:ou-process}
    d X_{t} = - \frac{1}{\tau} X_{t} dt + \epsilon \sqrt{\frac{2}{\tau}} dW_{t},
\end{equation}
where $\tau$ and $\epsilon$ are constants, and $W_{t}$ is the Wiener process. When written in this form, its asymptotic covariance for $s,t \gg 1$ is
\begin{equation}
    \langle X_{s} X_{t} \rangle = \epsilon^{2} e^{- |t-s| / \tau}.
\end{equation}
Therefore, one can use Eq.~(\ref{eq:ou-process}) to simulate Eq.~(\ref{eq:ebcm-lm}) by replacing $\tau$ and $\epsilon$ with their $l$-dependent counterparts. Alternatively, comparing the covariance of the forcing term in Eq.~(2) with that of a white noise process, $\langle \dot{W}(t) \dot{W}(t') \rangle = \delta(t-t')$, one can also identify Eq.~(\ref{eq:ebcm-lm}) as an OU process with $S_{lm} = (2\tau_{l})^{1/2} \epsilon_{0} \dot{W}$.

Numerically, the OU process in Eq.~(\ref{eq:ou-process}) can be simulated as follows:
\begin{equation} \label{eq:ou-discretization}
    X_{i+1} = e^{-\Delta t / \tau} \, X_{i} + \sqrt{\epsilon^{2}\left(1 - e^{-2 \Delta t / \tau}\right)} \, \Theta_{i}, 
\end{equation}
where $\Theta_{i} \sim N(0,1)$, drawn independently at each step. Here, Eq.~(\ref{eq:ou-discretization}) was advanced starting from the asymptotic variance, i.e., $X_{0} \sim N(0, \epsilon_{l}^{2})$, to sample immediately from the statistical steady state. The time step, $\Delta t = 0.5$~days, was taken to match the observed sample rate. As the fields are statistically isotropic, independent realizations were generated for each $m$.

The resulting angular and temporal power spectra of the realization used in Fig.~4 are shown in Fig. \ref{fig:figs01} (orange). This figure confirms that the realized background follows the $\mathcal{E}$BCM with the estimated parameters (black). In addition, it provides a sense of the effect of spectral leakage associated with the analysis and finite sampling (the upward inflection at $\omega > 0.6$ cpd).

\subsubsection{Statistical analysis \label{si:statistical-analysis}}
In the course of comparing the observed OLR with a realization of background, we wish to find regions (in either grid or spectral space) where the variance of the two differ. To this end, we use bootstrapping. This approach does not assume a priori that the samples are normally distributed and does not require an estimation of the number of degrees of freedom. In the following, we describe the analysis in physical space, used to compare the observed OLR variance with the background realization in Fig. 1 of the main text. The same analysis was also used in spectral space to compare the PSDs of the observed OLR and the background realization in Fig.~4 (C,F) of the main text. 

Let $S^{2}_{\mathrm{OLR}}$ and $S^{2}_{\mathcal{E}\mathrm{BCM}}$ denote the sample variance of the observed OLR and the background realization, respectively. The null hypothesis is $S^{2}_{\mathrm{OLR}} = S^{2}_{\mathcal{E}\mathrm{BCM}}$. As explained in the text, regions of suppressed variance are just as important as regions of enhanced variance. Therefore, the alternative hypothesis is $S^{2}_{\mathrm{OLR}} \neq S^{2}_{\mathcal{E}\mathrm{BCM}}$. The test statistic is $S^{2}_{\mathrm{OLR}} / S^{2}_{\mathcal{E}\mathrm{BCM}}$, and the significance level is 0.001.

Recall that the data were divided into 59 windows of length 360~days. Therefore, at each point on the globe, we have 59 samples from the OLR record and 59 samples from the background realization. These samples were combined to form a series of length 118, from which 5000 bootstrap samples (of length 118 each) were generated by resampling uniformly with replacement. Next, for each bootstrap sample, the ratio $S^{2}_{1} / S^{2}_{2}$ was calculated, where $S^{2}_{1}$ and $S^{2}_{2}$ are the sample variance of the first and last 59 readings. As the choice of the numerator/denominator is arbitrary, the p-value was calculated as the fraction of samples for which this ratio is greater than the test statistic or smaller than its reciprocal. This process was repeated for each point on the globe.

\begin{figure*}
	\includegraphics[width=0.9\textwidth]{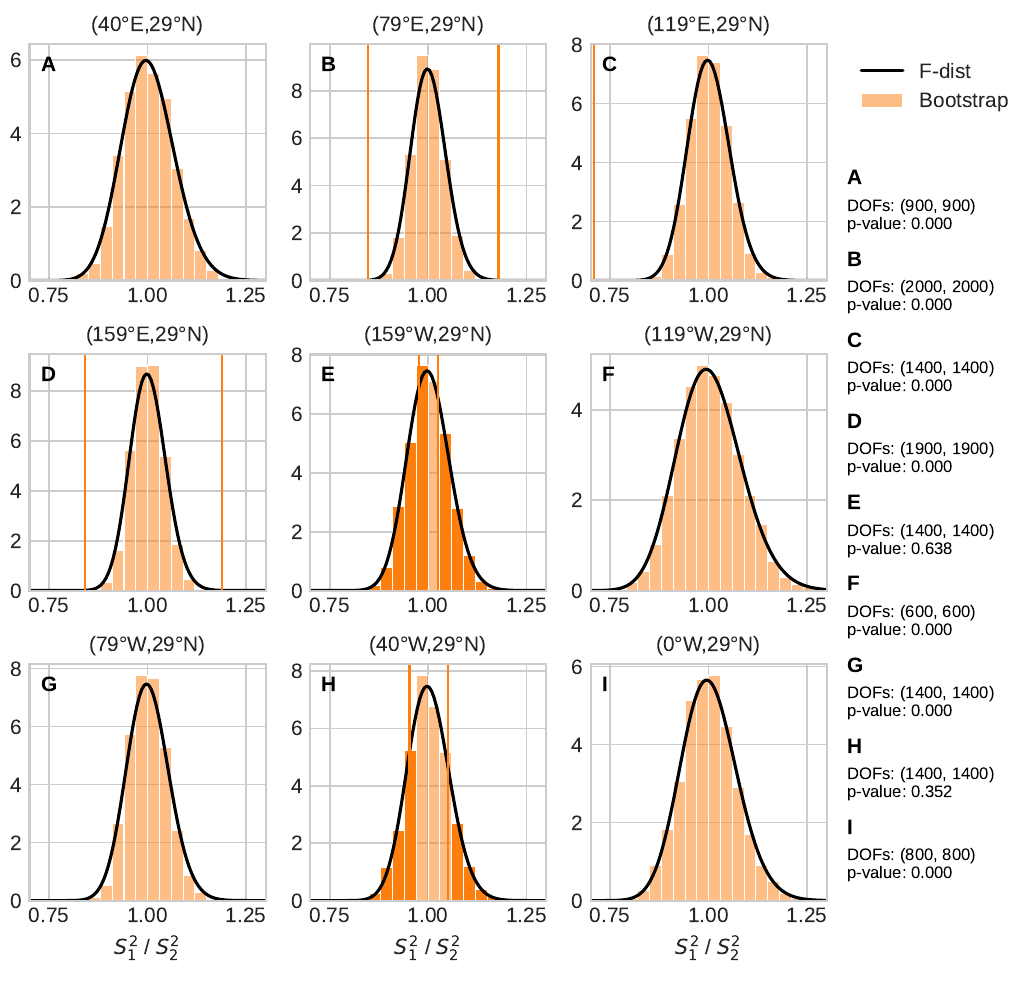}
	\caption{\label{fig:figs02}\textbf{Bootstrap statistics.}
		The Probability Distribution Functions (PDFs) of the bootstrap samples (light orange) at 9 representative points. The test statistic and its reciprocal are marked by vertical orange lines, and the more extreme values found in the bootstrap samples are emphasized (dark orange). Except for panels (B, D, E, H), the test statistic and its reciprocal are outside of the shown domain.}
\end{figure*}

The Probability Distribution Functions (PDFs) of the bootstrap samples (light orange) are shown in Fig.~\ref{fig:figs02} for 9 representative points. The test statistic and its reciprocal are marked by vertical orange lines, and the more extreme values found in the bootstrap samples are emphasized (dark orange). Except for panels (B, D, E, H), the test statistic and its reciprocal are outside of the shown domain. In particular, except for panels (E, H), the p-value is less than 0.001, and the probability of drawing more extreme values than the test statistic is negligible. In general, we find that the p-values are less than 0.001 throughout most of the globe, except regions of transition between high and low variability (Fig. 1 of the main text). The PDFs of the bootstrap samples follow an F-distribution (black line), albeit with widely different and unpredictable degrees of freedom. This finding provides reassurance that the chosen method is appropriate for comparing the variance of the observed OLR and background realization, and raises questions about methods that require an estimate of the number of degrees of freedom.

\subsection{SI 2: Additional details on the stochastic energy balance climate model ($\mathcal{E}$BCM) \label{si:ebcm}}

The stochastic $\mathcal{E}$BCM used here to describe the Earth's infrared background is based on the one studied in~\citet{north1981predictability, kim1991surface,north2011correlation}. The key difference is in the details of the stochastic forcing. Unlike these works and unlike~\citet{hottovy2015spatiotemporal}, we find it necessary to allow the forcing to be scale-dependent to explain the observed OLR variance while adhering to the fluctuation-dissipation theorem (FDT).

Using the fact that the Spherical Harmonics are the eigenfunctions of the Laplacian in spherical coordinates, with eigenvalues $-l(l+1) / a^{2}$, the spectral space projection of Eq.~(1) is a standard Langevin equation for each $l$ and $m$ (independent of $m$), i.e.
\begin{equation} \label{eq:ebcm-lm}
    \frac{d F_{lm}}{dt} = - \frac{1}{\tau_{l}} F_{lm} + \frac{1}{\tau_{0}} S_{lm},
\end{equation}
where $\tau_{l} = \tau_{0} / [1 + \lambda_{0}^2 l(l+1) / a^{2}]$. Recall that we take the forcing to be Gaussian white noise in time and statistically isotropic in space. For the sake of exposition, in this section, we represent the forcing heuristically as the derivative of a (non-differentiable) Wiener process, writing
\begin{equation} \label{eq:ebcm-lm-wiener}
    \frac{d F_{lm}}{dt} = - \frac{1}{\tau_{l}} F_{lm} + \frac{s_{l}}{\tau_{0}} \dot{W}.
\end{equation}
Here, we have used the fact that the forcing is isotropic in space, so that the Wiener process variance $s_{l}^{2}$ depends only on $l$. A more rigorous analysis can be carried out using only the second moment of the forcing, as in Eq.(2) of the main text, and following the analysis in~\citet{kubo1966fluctuation}.

For any choice of $s_{l}$ and $\tau_{l} > 0$, Eq.~(\ref{eq:ebcm-lm-wiener}) describes an Ornstein--Uhlenbeck (OU) process with equilibrium variance
\begin{equation} \label{eq:ou-stationary-variance}
    \langle F_{lm}^{} F_{lm}^{*} \rangle_{\text{eq}} = \frac{s_{l}^{2} \tau_{l}}{2 \tau_{0}^{2}}.
\end{equation}
The FDT effectively inverts this relation, implying that the forcing variance $s_{l}^{2}$ is determined by the equilibrium variance of the system. In classical Brownian motion, the latter can be further related to the ambient temperature from first principles via the Maxwell-Boltzmann distribution or the equipartition theorem. In the present work, the equilibrium variance can only be estimated from observations. Nevertheless, except for the special case where $\langle F_{lm}^{} F_{lm}^{*} \rangle_{\text{eq}}  \propto \tau_{l}$, the forcing variance is scale-dependent. 

In the text, the equilibrium variance is denoted by $\epsilon_{l}^{2} = \langle F_{lm}^{} F_{lm}^{*} \rangle_{\text{eq}} $, and, for convenience, we have also defined $\epsilon_{0}$ such that $\epsilon_{l} = \epsilon_{0} \tau_{l} / \tau_{0}$. With these definitions, the forcing variance is
\begin{equation}
    s_{l}^{2} = 2 \epsilon_{0}^{2} \tau_{l},
\end{equation}
as in Eq.(2). Finally, the process in Eq.~(\ref{eq:ebcm-lm-wiener}) is fully determined by its covariance function, yielding Eq.~(3)~\cite[e.g.,][]{zwanzig2001nonequilibrium}. The PSD in Eq.~(4) is related to the covariance via the Wiener–Khinchin theorem, i.e., $\hat{C}_{l}(\omega) = \int_{-\infty}^{\infty} C_{l}(\tau) e^{-i\omega\tau} d\tau$.

\subsection{S3: Supporting results}

\subsubsection{Spatiotemporal correlations \label{si:correlations}}

The spatial de/correlation in the $\mathcal{E}$BCM is not described by an e-folding scale (it is not an AR-1 or an OU process in space). Instead, as is shown in~\citet{north2011correlation}, the spatial covariance, in frequency space, between two points on the sphere separated by a central angle $\theta$ apart, is
\begin{equation} \label{eq:spatial-correlation}
\hat{C}_{\text{Response}}(\cos\theta, \omega) = \sum_{l=0}^{\infty} \frac{(2l+1)}{4\pi} \hat{C}_{l}(\omega) \, P_{l}(\cos\theta),
\end{equation}
where $\hat{C}_{l}(\omega)$ is the PSD given by Eq.~(4) of the main text, and $P_{l}(\cos\theta)$ are the Legendre polynomials of degree $l$. The analysis in~\citet{north2011correlation} can also be used, with straightforward modifications, to compute the spatial correlation associated with the forcing, yielding
\begin{equation} \label{eq:spatial-correlation-forcing}
\hat{C}_{\text{Forcing}}(\cos\theta) = \sum_{l=0}^{\infty} \frac{(2l+1)}{4\pi} 2\epsilon_{0}^{2}\tau_{l} \, P_{l}(\cos\theta).
\end{equation}

\begin{figure*}
	\includegraphics[width=12cm]{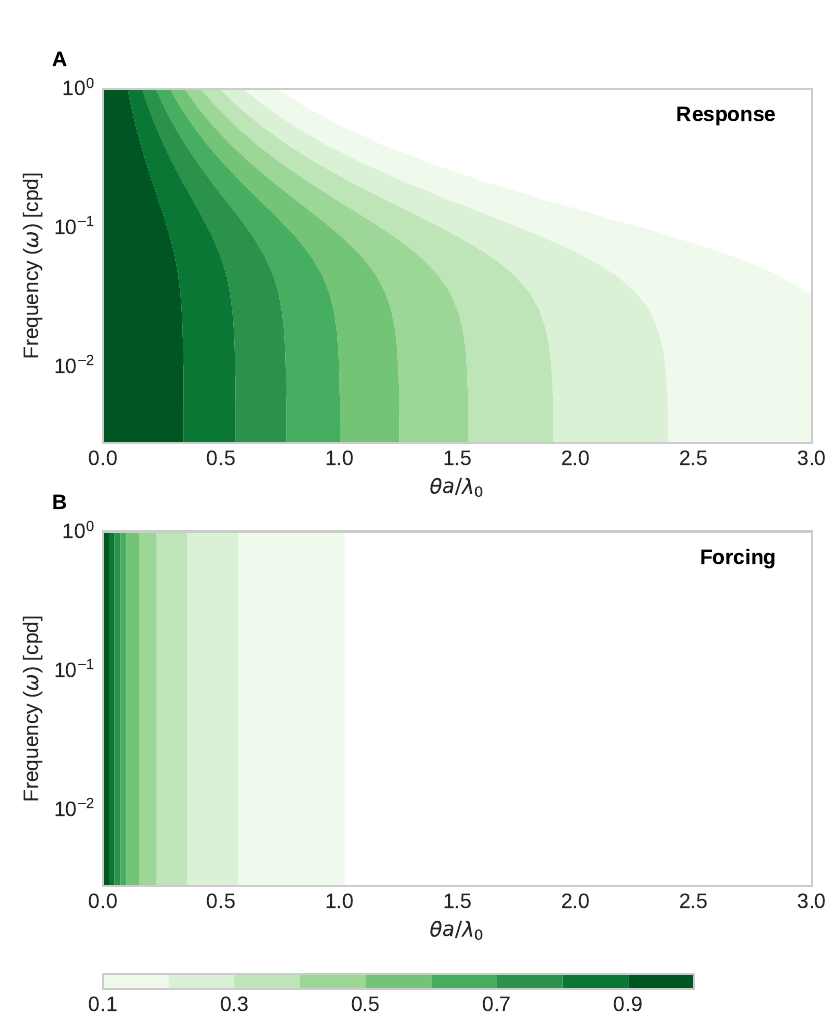}
	\caption{\label{fig:figs03}\textbf{Spatial correlation.}
		Contour plots of the spatial correlation as a function of the frequency on the ordinate and the geodesic distance on the abscissa. The latter is scaled on $\lambda_{0}$, so that a value of $\theta a / \lambda_{0} = 1$ corresponds to one decorrelation length, about 400~km for the observed infrared background. (A) The spatial correlation of the response implied by Eq.~(\ref{eq:spatial-correlation}). (B) The spatial correlation of the forcing implied by Eq.~(\ref{eq:spatial-correlation-forcing}).}
\end{figure*}

\sloppy Fig. \ref{fig:figs03}A shows a contour plot of the correlation implied by Eq.~(\ref{eq:spatial-correlation}), i.e. $\hat{C}_{\text{Response}}(\cos\theta) / \hat{C}_{\text{Response}}(1)$, as a function of the frequency on the ordinate and the geodesic distance on the abscissa. The latter is scaled on $\lambda_{0}$, so that a value of $\theta a / \lambda_{0} = 1$ corresponds to one decorrelation length, about 400~km for the observed infrared background. For low frequencies, the correlation after one decorrelation length is still substantial, about 0.6. Even after three decorrelation lengths ($\theta a / \lambda_{0} = 3$) the correlation is non-negligible, about 0.1. In other words, a correlation of 0.1 between two points distanced 1200~km apart can simply be associated with long-period random noise. As the frequency increases, the correlation at a given distance decreases. In other words, faster fluctuations decorrelate faster. In particular, for $\omega = 1$ cpd, which represents our Nyquist frequency, the correlation after one decorrelation length is negligible. For comparison, the correlation associated with the forcing is shown in panel (B). For low frequencies, the forcing decorrelates faster than the response at any given distance, whereas, for high frequencies, the forcing decorrelates faster than the response only in the vicinity of $\theta = 0$. Still, even at high frequencies the two are well separated.

\subsubsection{Sensitivity of the angular variance to averaging over $m$ \label{si:m-sensitivity}}

\begin{figure*}
	\includegraphics[width=0.9\textwidth]{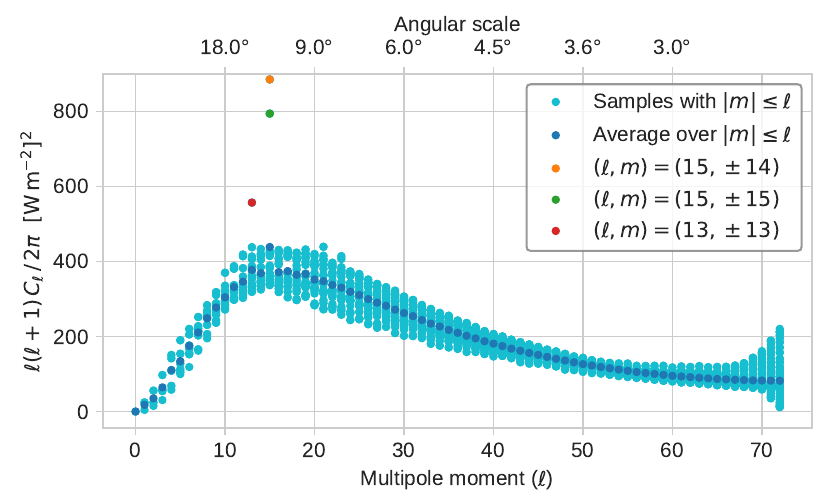}
	\caption{\label{fig:figs04}\textbf{Sensitivity of the angular variance to averaging over $m$}. A scatter plot of $C_{lm} = \langle F_{lm}^{} F_{lm}^{*} \rangle$ as a function of $l$, for all $|m| \leq l$ (cyan), compared to the angular variance ${C}_{l} = \sum_{m=-l}^{l} \langle F_{lm}^{} F_{lm}^{*} \rangle \, / (2l+1)$ (blue). The former is estimated by averaging the spectral space OLR anomalies over time. The latter, also shown in Fig.~2 of the main text, is obtained by further averaging $C_{lm}$ over $m$. The outliers at $(l, m) = (15, \pm 14)$ (orange), $(15, \pm 15)$ (green), and $(13, \pm 13)$ (red) are colored differently to emphasize that the outliers in Fig.~2 of the main text result only from these particular values of $m$.}
\end{figure*}

The angular variance in Fig.~2 of the main text was obtained by averaging the spectral space OLR fluctuations $F_{lm}^{}(t) F_{lm}^{*}(t)$ over time and $m$ (with $|m| \leq l$) to estimate the angular variance ${C}_{l} = \langle F_{lm}^{} F_{lm}^{*} \rangle$. Fig. \ref{fig:figs04} shows a scatter plot of $C_{lm}$, without averaging over $m$, as a function of $l$, for all $|m| \leq l$ (cyan), compared to $C_{l}$ (blue, same as Fig.~2 of the main text). The spread associated with $m$ is larger than that implied by the standard error in Fig.~2 of the main text. However, the latter is greatly reduced by averaging over both $m$ and the temporal windows. In addition, the spread in $m$ is generally distributed around the variance predicted by the $\mathcal{E}$BCM. This figure also shows that the outlier at $l=15$ results only from $m = \pm 14$ (orange) and $m = \pm 15$ (green). The former is consistent with the number of swaths seen by the satellites per day \cite{wheeler1999convectively}. The latter may be excited by the former, but is otherwise unexplained. Likewise for the outlier at $l=13$ and $m = \pm 13$ (red). 

\subsubsection{Spectrally filtered spatial variance \label{si:spectral-filters}}

The spectral filters in Fig.1 of the main text include only the two branches of eastward and westward non-dispersive variability, as well as the equatorial Kelvin filter for comparison. The global background highlights 3 more distinct regions of significance in spectral space. Therefore, to complete the picture, we provide the remaining filters in this section. 

Fig. S5 provides a close up on Fig. 4F of the main text, and the filtered spatial variance is shown in Fig. S6. The filters were defined somewhat heuristically, but the results remain qualitatively similar with small changes in these filters. As in the main text, the filters were applied to both the observed OLR and the realized background, and their spatial variance were subtracted. As mentioned in the text, the spatial variance in the central lobe (Filter 4) consists of a zonally symmetric equator-to-pole gradient, highlighting the lower OLR variability at the poles observed in Fig. 1A. The spatial variance of the side lobes (Filters 5 and 6) highlights regional variability in the monsoonal regions, including the North American, South American, North African, South African, Indian, East Asian, and South Asian Monsoons. They also highlight the ITCZ the storm track regions. Like the non-dispersive branches, the latter is suppressed in the westward lobe, consistent with the extratropical jets. In contrast to the non-dispersive branches, the western lobe has more variance. 

Finally, we also include here the MJO and equatorial Rossby wave filters, for comparison, even though these features are not significantly different from the background in spectral space. It is evident that the spatial patterns of the westward non-dispersive branch and the western side lobe are similar to that of the equatorial Rossby wave, albeit the former is more squished around the equator. 

\begin{figure*}
	\includegraphics{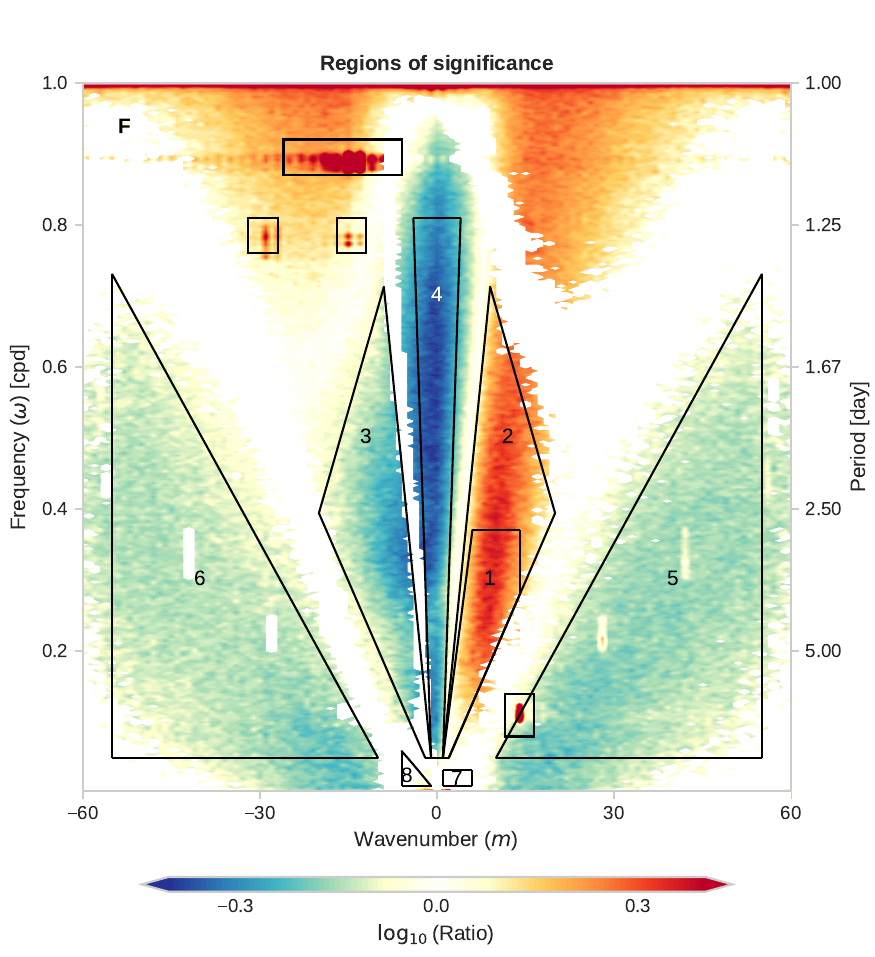}
	\caption{\label{fig:figs05}\textbf{Spectral filter definitions}. A close up on panel \textbf{F} of Fig. 4 in the main text. The wavenumber range from -60 to 60. For comparison, the MJO and equatorial Rossby wave filters (labeled 7 and 8) are included in the Supporting Information, despite not being significantly different from the background. For simplicity, the lower limit of the latter is taken to be constant and equal to 1/96~cpd, rather than the dispersion curve with an equivalent depth of 8~m as in~\cite{wheeler1999convectively}.}
\end{figure*}

\begin{figure*}
	\includegraphics[width=16cm]{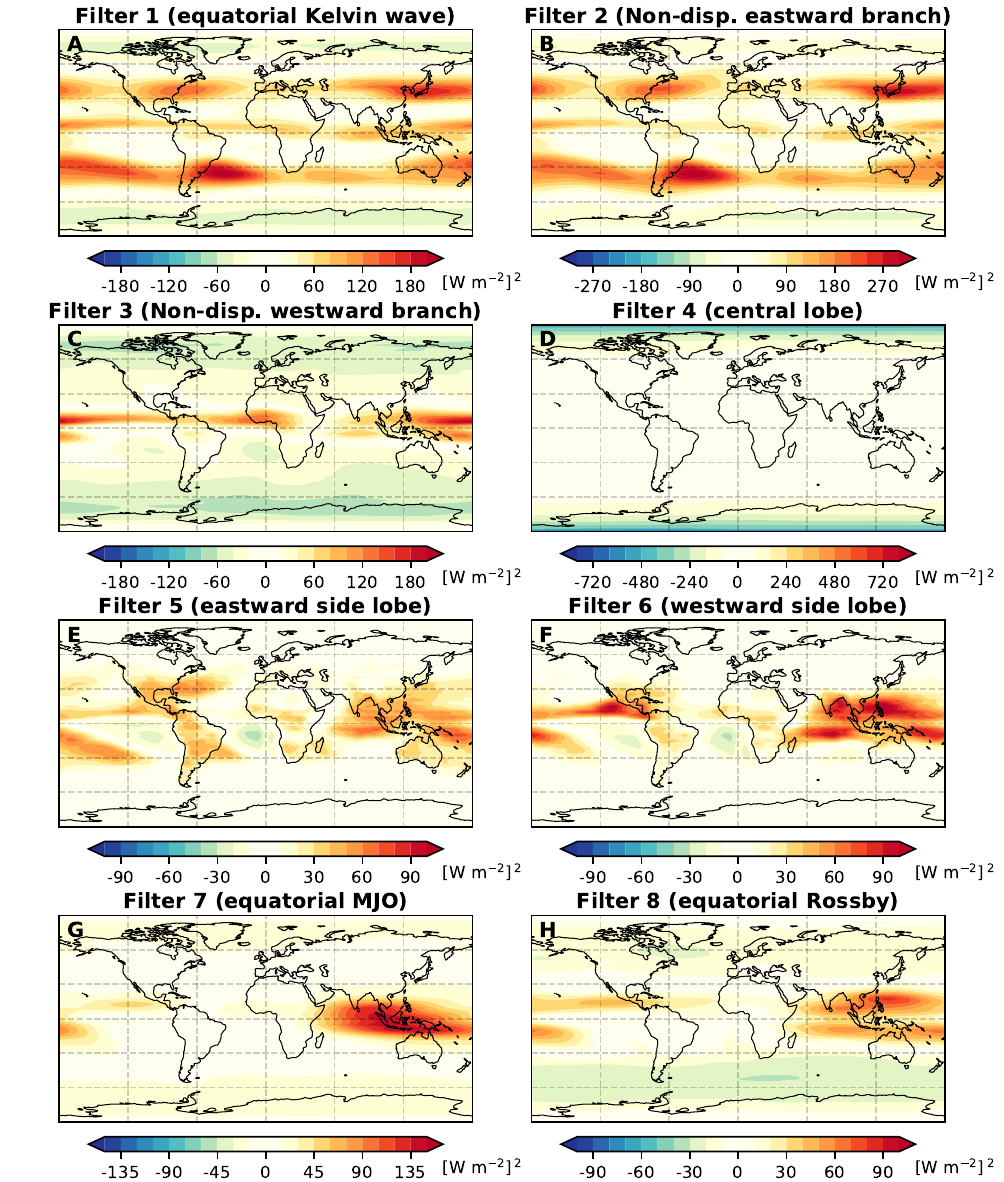}
	\caption{\label{fig:figs06}\textbf{Spectrally filtered spatial variance}. Same as Fig.1 of the main text, but for the eight filters outlined and labeled in Fig. S5. Panels A, B, C are the same as panels B, C, D of Fig. 1 of the main text, respectively.}
\end{figure*}

%

\bibliography{references}

\end{document}